\documentstyle[osa,manuscript,epsfig]{revtex}

\newcommand{\tj}{$t$-$J$\ }

\newcommand{\etal}{{\it et.~al.}}
\newcommand{\ij}{\langle ij\rangle}

\begin{document}


\title{Why do stripes form in doped antiferromagnets\\
and what is their relationship to superconductivity?}
\author{ Steven R.\ White$^1$ and D.J.\ Scalapino$^2$}
\address{ 
$^1$Department of Physics and Astronomy,
University of California,
Irvine, CA 92697
}
\address{ 
$^2$Department of Physics,
University of California,
Santa Barbara, CA 93106
}
\maketitle
\begin{abstract}
\noindent 

Experiments show evidence for stripe formation in the
underdoped cuprates.  Here we discuss recent numerical
calculations on the \tj model which tell us
about the mechanism responsible for stripe formation and
the relationship between stripes and superconductivity.

\end{abstract}


It is clear from a variety of experiments \cite{stripeexper} 
that stripes appear as
important low-energy configurations in the underdoped cuprates.
However, the basic questions of why stripes form and what role
they play in superconductivity remain controversial. A decade
ago Hartree-Fock solutions of the Hubbard model showed that
stripes---one-dimensional domains of increased hole density
forming anti-phase N\'eel boundaries---were present in mean field
solutions of the Hubbard model \cite{hfdomain}.  
Here the stability of the
stripe structure arises from the reduction in kinetic energy
that the holes experience in moving transverse to the stripes. 
However, the stripes in the Hartree-Fock solution are
characterized by a filling of one hole per domain wall unit
cell, while experiments on the cuprates at low doping find a
filling of half this.  In addition, within the Hartree-Fock
framework, it is not clear how superconductivity enters. 

An alternative view argues that stripe formation arises from
competition between phase separation and long-range Coulomb
interactions \cite{emery}. 
Central to this ``frustrated phase separation''
picture is the assumption that lightly doped $t-J$ or
Hubbard models, with parameters in the relevant physical regime,
will, in the absence of a long-range Coulomb interaction, globally phase
separate into uniform hole-rich and undoped regions.  In this
approach, it is argued that the formation of stripes is governed
by a larger charge energy scale and that pairing arises as a secondary
effect associated with the transfer of a spin gap from the
undoped regions to the stripes and a subsequent pair transfer
between stripes, leading to a Josephson coupling and
superconductivity. The fact that stripes act as
antiferromagnetic domain walls is also a secondary ``kinetic''
effect in this picture.

Recently we have carried out numerical density matrix
renormalization group (DMRG) calculations \cite{dmrg} on various $t-J$
systems which suggest a third view.  These calculations show
that low-lying striped states occur in the \tj model in the
absence of long-range Coulomb interactions\cite{stripe,energetics}.
Furthermore, unlike the Hartree-Fock solutions, the domain
walls are characterized at low doping by a filling of one hole
per two domain wall unit cells. The short-range
structure of the domain wall contains strong antiferromagnetic
singlet bond correlations crossing the holes. Just as in the
case of the two-hole bound state \cite{holestructures}, these
spin correlations around and across the holes form in order to
maximize the hopping overlap with other hole configurations,
which lowers the kinetic energy, while at the same time
minimizing the disturbance of the AF background. For this reason
it is not surprising that the binding energy per hole of a
domain wall is only slightly greater than that of a hole in a
$d_{x^2-y^2}$ pair.
When an additional next-nearest-neighbor, one-electron hopping
term is added, the tendency to form stripes is weakened\cite{Toh99}  and 
we find that the domain walls become unstable
with respect to the $d_{x^2-y^2}$ pairing state \cite{tprime}.
These calculations suggest:
\begin{enumerate} 
\item When two holes are added to the undoped system, a pair forms as the holes
locally arrange themselves so as to satisfy the
competing requirements of minimizing their kinetic energy and
minimizing the disturbance of the background exchange interactions. 
At finite doping, domain walls form for similar local reasons
and to support $\pi$
phase-shifted regions on either side 
as a way of further reducing the disturbance of
the exchange interactions and to lower the transverse kinetic
energy of the holes. 

\item The domain walls have a minimum energy for a linear filling of
$\rho_\ell=0.5$, corresponding to one-hole per two domain-wall unit cells. 
Domain walls form at this linear filling for hole dopings $x < 1/8$,
at which point the repulsion between domain walls becomes large
enough so that additional holes cause walls with $\rho_\ell > 0.5$ to form.

\item Stripes {\it compete} with superconductivity, and by changing the
parameters of the system, a pairing state can be obtained as the
stripes {\it evaporate}. In particular, a change in parameters
which enhances pair mobility can destabilize the stripes,
leading to a stronger pairing state.

\end{enumerate}
\noindent
In the following we will discuss numerical results for the \tj
model which lead us to this view. 

The Hamiltonian of the basic \tj model is
\begin{equation}
H = - t \sum_{\langle ij \rangle s}
      ( c_{is}^{\dagger}c_{js}
                + {\rm h.c.}) + 
J \sum_{\langle ij \rangle}
      ( {\bf S}_{i} \! \cdot \! {\bf S}_{j} -
         \frac{n_i n_j}{4} ) ,
\label{tj-ham}
\end{equation}
where doubly occupied sites are explicitly excluded from the Hilbert
space.
Here $\ij$ are nearest-neighbor sites, and $s$ is a spin index. The operator 
and $c^\dagger_{is}$ creates an electron of spin $s$ on site $i$ and $\vec
S_i = \frac{1}{2}\, c^\dagger_{is} \vec\sigma_{ss^\prime} c_{is^\prime}$ and
$n_i=c^\dagger_{i\uparrow} c_{i\uparrow} + c^\dagger_{i\downarrow}
c_{i\downarrow}$ are the electron spin moment and charge density operators
at site $i$.
The nearest-neighbor hopping and exchange interactions are $t$ and $J$,
and the average site occupation $<n>=1-x$ is set by the hole
doping parameter $x$.

In the absence of hole doping ($x=0$) the \tj model reduces to a
Heisenberg $S=1/2$ antiferromagnet and the 2D ground state has
long range antiferromagnetic order\cite{reger}. The question is,
what happens when it is doped? At sufficiently large values of
$J/t$, the doped system will globally phase separate into a
hole-free antiferromagnetic region and a hole rich region.
However, in the physically relevant regimes of $J/t$ and doping
discussed here, the system does not exhibit global phase
separation\cite{comparison}. Rather, the DMRG
calculations for the \tj model which we will discuss here have 
striped ground states.  On various hole-doped \tj lattices we have 
observed bond-centered
and site-centered domain walls which run along the (1,0) or (0,1) direction
as well as diagonal domain walls which run along the (1,1) direction,
depending upon the nature of the boundary conditions. In all cases, these
domain walls separate $\pi$-phase shifted AF regions, and exhibit
antiferromagnetic bonds crossing the holes. The (1,0) and (0,1)
walls have a lower energy with the bond-centered and
site-centered walls very close in energy. Figure 1(a) shows a
typical bond-centered (0,1) domain wall structure.
Here the charge density
and spin structure on a centered $8\times 8$ section of a $16\times 8$
lattice is shown. For the results shown in Fig. 1, $J/t=0.35$, $x=0.125$,
and the lattice has periodic boundary conditions in the
$y$-direction and open boundary conditions in the $x$-direction
with a staggered magnetic field of strength $h=0.1t$ applied to
the open ends. On the full $16\times 8$ lattice
there are four vertical bond-centered stripes separating $\pi$-phase
shifted antiferromagnetic regions. Two of these stripes on the central
$8\times 8$ section of the lattice are shown in Fig. 1(a). Each of the
stripes contains 4 holes corresponding to a linear filling of 1 hole per
two-domain wall unit cells. Fig. 2(a) shows a $13\times8$
lattice with 12 holes and with a $\pi$-phase shift in the
staggered magnetic field which is applied at the ends. In this
case, domain walls form which are site centered. Just as
for the bond-centered domains, the site-centered domains shown
in Fig. 2(a) have a linear filling of one hole per two domain wall
unit cells.

To understand why such stripes form in the \tj model, it is useful to examine
these domain walls in more detail.
In Fig.~1(b), we show results from a smaller system, of size
$8\times8$, with the same hole doping and boundary conditions as
in (a). The spin and hole densities for this system (not shown)
are almost identical to
those shown in (a). The blue circles show the most probable
configuration of the eight holes in the system.  The thickness of the lines
connecting various sites denotes the strength of the exchange field
$\langle \vec S_i \cdot \vec S_j\rangle$ for this configuration of the holes. 
The maximum strength of this exchange field corresponds to a singlet bond
with $\langle\vec S_i \cdot \vec S_j\rangle=-3/4$.
A similar construction for the site-centered case is shown in
Fig. 2(b).
Note the numerous frustrating valence bonds crossing the holes,
most of which connect opposite sides of each domain wall. In
fact, we find that a tendency for frustrating valence bonds to
form across mobile holes is universal in doped
antiferromagnets, coming directly from the local competition
between the kinetic and exchange energies\cite{holestructures}. 
A domain wall allows most of these exchange bonds to form in a way
which cooperates, rather than competes, with the background
spin configuration.
The configurations shown for these two cases in Figures 1(b) and 2(b) 
are, of course, only one 
out of the huge number which form
the ground states, but they provide a local strong coupling 
picture of the type of correlations which
are present in the domain walls. 

It is interesting to compare the domain wall  
configuration of Fig.~1(b) with the two most probable configurations of 
two holes on an $8\times 8$ lattice shown in Fig.~3(a). 
In the unphysical regime $J>t$, the most probable configuration 
for the two holes
would tend to be near neighbors to reduce the number of broken
exchange bonds. In the physical region, $J<t$, the kinetic
energy plays an increasingly important role so that the diagonal
configuration of the holes shown on the left-hand side of
Fig.~3(a) is the most probable. When the holes sit on diagonal
sites, there is also a large diagonal singlet correlation 
\cite{holestructures}. 
As shown in Fig.~3(b), four of the eight one-electron hops from this diagonal
configuration lead to a configuration in which the diagonal
singlet becomes a nearest-neighbor exchange bond. The
phasing of the near neighbor hole configurations has
$d_{x^2-y^2}$ symmetry as does the diagonal site configuration
when the background spins are taken into
account\cite{holestructures,poilblanc}. Thus, in this
strong coupling picture the pairing arises from a compromise in
which the holes locally arrange themselves so as to minimize the
disturbance of the background exchange energy while at the same
time lowering their kinetic energy by the hopping between
configurations such as those shown in Fig.~3(b). 

In a similar way, the site-centered domain wall configuration of Fig.~2(b)
reminds one of the valence bond-like region that forms around holes in the
one-dimensional $t$-$J$ model \cite{holestructures,MED99}. Again, to
maximize the hopping overlap with adjacent hole configurations one expects
there to be strong antiferromagnetic correlations between
next-nearest-neighbor sites across a hole.  Such a valence bond-like
correlation becomes a nearest-neighbor link after one hop of the hole to
either neighboring site, since moving the hole also moves the bond. 

We see that the domain walls {\it locally} 
share a number of the same features as the two-hole pairing state,
which accounts for the fact that the energy per hole for a
domain wall is close to the energy per hole of a pair. In
addition, however, by lining up to form a domain wall, the holes can
support a
$\pi$-phase shift in the surrounding antiferromagnetic
background, lowering the energy
further. For the site-centered wall this reflects the reduction in the
kinetic energy that holes experience in moving transverse to the domain
walls across which there are strong antiferromagnetic bonds (see Fig.~2(b).
A similar effect occurs for the bond-centered  domain wall shown in Fig.~1.
In this case it is useful to think of local pairs lined up to support the
$\pi$-phase-shift in the surrounding antiferromagnitc background. The
stabilization energy of a $\pi$-phase shifted field on a pair
can be calculated for the simple case of two holes on
$2\times 2$ \tj lattice, such as that shown in Fig.~3(b). Representing the
exchange field which runs along both sides of the stripe by a mean field of
magnitude $h$, and numbering the site counterclockwise around the
plaquette, the perturbation added to the \tj Hamiltonian for the
$\pi$-phase-shifted antiferromagnetic domain wall is 
\begin{equation}
H_{\pi AF} = h\left(S^z_1 + S^z_2 - S^z_3 - S^z_4\right)
\label{two}
\end{equation}
while if there were no phase shift
\begin{equation}
H_{AF} = h\left(S^z_1 - S^z_2 + S^z_3 - S^z_4\right)\ .
\label{three}
\end{equation}
In second order, one finds that either perturbation  lowers the
energy per hole by a term of order $h^2/t$ when $J/t$ is small,
but the coefficient in the $\pi$-phase-shifted field case is
about four times larger. The $\pi$-phase-shifted
antiferromagnetic
field at the edges of the domain mixes in a spin-triplet
contribution to the local pairing correlations, lowering both
the exchange and kinetic energies.
This is similar to the
effect that Krotov et.~al. discussed\cite{krotov}, 
for the case of a 2-leg ladder,
within a weak coupling
renormalization group approach.  There, because they neglected
Umklapp processes, which may lead to a charge density wave
(CDW) for
$\rho_\ell=0.5$, they found that a $\pi$-phase-shift field on the edges of a
two-leg ladder enhanced the pairing correlations. Here we find that the
energy per hole of such a domain wall is lowered, but there are only short
range pairing correlations.

In order to examine the properties of a domain wall in more detail, we have
studied a single long domain wall which forms down the center of
$16\times4$ and $16\times 6$ lattices which have 
a $\pi$-phase shifted staggered magnetic
field $0.1t$ applied to the top and bottom edges.
The energy per hole for this
system defined relative to an undoped ladder of the same size with $AF$ edge
boundary conditions is shown in
Fig.~4 versus the linear domain wall filling $\rho_\ell=N_{\rm holes}/L$
with $L=16$. One sees that the energy is a minimum on the $16\times 6$
lattice for a linear filling of one
hole per two-domain wall unit cells, $\rho_\ell=0.5$, and is concave for this
size system for $0.5<\rho_\ell<1.0$. On a longer $40\times 6$
lattice we have seen phase separation into segments with $\rho_\ell \simeq
1.0$ and $\rho_\ell\simeq 0.5$ when the average filling is between 0.5 and 1.0.
Similar results are found for the site centered domain wall.
Additional simulations have shown that domain walls repel each other at
short distances so that at low doping ($x<1/8$) one has an array of
half-filled domain walls with a spacing $d=(2x)^{-1}$. At $x=1/8$, the
repeat distance for the walls is four lattice spacings as seen in Fig.~1.
At fillings greater than 1/8, the wall repulsion is sufficiently strong
that, rather than reduce the wall spacing further, some domain walls with
$\rho_\ell=1.0$ are added leading to a change in the $d$ versus $x$ relation
at $x=1/8$.

We have also measured the pair field correlations along the $\rho_\ell=0.5$
domain wall in the $16\times 4$ lattice.
Fig.~5 shows the pair-field
correlations along the central two legs with
\begin{equation}
D_{yy}(\ell) = \left\langle\Delta_y(i+\ell) \Delta^\dagger_y
(i)\right\rangle
\end{equation}
and
\begin{equation}
D_{xy}(\ell) = \left\langle\Delta_x (i+\ell) \Delta^\dagger_y
(i)\right\rangle
\label{four}
\end{equation}
Here
\begin{equation}
\Delta^\dagger_y(i) = c^\dagger_{i,2\uparrow}c^\dagger_{i,3\downarrow}-
c^\dagger_{i,2\downarrow} c^\dagger_{i,3\uparrow}
\label{five}
\end{equation}
is an operator that creates a singlet pair on the $i^{\rm th}$ rung between
leg 2 and leg 3, and
\begin{equation}
\Delta_x(i) = c_{i+1,2\downarrow} c_{i,2\uparrow} - c_{i,2\downarrow}
c_{i+1,2\uparrow}
\label{six}
\end{equation}
destroys a singlet pair on leg 2 between the $i$ and $i+1$ rungs. The short
range $d_{x^2-y^2}$-like structure of the pairing correlations is seen in
the sign change between $D_{yy}(\ell)$ and $D_{xy}(\ell)$. The
pair-field correlations are clearly suppressed at larger distances. 

The suppression of pairing along a domain wall can be understood
as arising from a suppression of charge fluctuations induced
by the $\pi$-shifted antiferromagnetic background. Strong local charge
fluctuations are essential for superconductivity. In a domain
wall with $\rho_\ell=0.5$, as two adjacent holes or hole pairs move away 
from each other,
the resulting region in which the two $\pi$-phase shifted domains are
in contact result in a restoring potential which grows
linearly with the separation. This strongly suppresses such charge
fluctuations, and leads to the decay of the longer-range pairing
correlations. Further suppression of pairing correlations may
come from a tendency for CDW formation at $\rho_\ell=0.5$.
We have found that a 2-leg ladder with a filling  $x=0.25$, 
corresponding to a linear filling $\rho_\ell=0.5$, 
has a small charge gap and long range CDW order in its ground state.

We have recently studied systems with a next nearest neighbor
hopping term $t'$ added to Eq.~(\ref{tj-ham}) \cite{tprime}. 
Although a variety of terms can be added to the basic \tj
Hamiltonian to improve its applicability to experimental systems, 
$t'$ is particularly interesting because it
directly affects the competition between pairing and stripe
formation.
Fig.~6(a) shows the hole and spin
density for different values of $t^\prime$ for a
$12\times 6$ system with periodic boundary conditions in the $y$-direction.
As $t^\prime$ increases, the static stripe structure is smeared out
\cite{Toh99} and, as
shown in Fig.~6(b), the pairing correlations are enhanced \cite{tprime}. 
We have measured
the density-density CDW correlations for the lattices with the smeared
out domain walls and find them to be negligible,
implying that the smearing out of the charge density is not due
to fluctuations of the domain walls, but rather a reduction in their ability to 
bind holes which eventually leads to the complete evaporation
of the stripes into pairs. 
The effect of $t^\prime$ is to enhance the pair mobility, leading
to a lowering of the stabilization energy of the domain walls. As this happens,
the pairing correlations increase and the stripes disappear. For $t^\prime
= 0.3t$ the antiferromagnetic response driven by the staggered
field at the open ends is peaked at $(\pi,\pi)$ and $d_{x^2-y^2}$-pairing
correlations are dominant. Fig.~6 clearly shows that the striped domain-wall
state and the superconducting pairing state compete for $t^\prime > 0$. 
However, there does appear to be an overlap region in which
pairing is significant but weakly bound domain walls remain. 

Thus, in the nearest-neighbor \tj model, domain walls are
energetically favored over pairs and we see only weak pairing
correlations. Turning on $t^\prime > 0$
enhances the pair mobility, tipping the balance towards a
$d_{x^2-y^2}$-pairing state. Phenomenologically, $t^\prime/t > 0$
models the electron-doped materials, with $x=\langle n_i\rangle -1$ the
electron doping rather than the hole doping. Thus, one might have expected
to see even stronger pairing correlations for $t^\prime/t < 0$,
corresponding to the hole doping case. However, as discussed in
\cite{tprime}, in this case the domain walls evaporate into
quasi-particles and the $d_{x^2-y^2}$-pairing correlations
remain weak\cite{noteaa}.

In summary, these results lead to the conclusions that in the \tj model,
stripes and pair formation are driven by the same basic mechanism, the
competition between kinetic and exchange energies, and that they compete
with each other. In the nearest-neighbor \tj model, domain-wall/stripe
formation is slightly favored over $d_{x^2-y^2}$ pairing.
At low doping the stripes are characterized by a linear
filling $\rho_\ell=0.5$ and the repulsion between the stripes give a stripe
spacing $d=(2x)^{-1}$. For a hole doping $x\ge 1/8$, some of the stripes
switch to a filling $\rho_\ell=1.0$ giving rise to a change of behavior at
this doping. In the nearest-neighbor \tj model, domain walls are
energetically favored over pairs, and the pairing correlations are weak. 
As the next-nearest-neighbor hopping $t^\prime$ is turned on, one goes
continuously from a situation in which the stripe correlations dominate to
one in which the pairing correlations are dominant. Similar
effects should be seen for other changes in the model which enhance
pair mobility or act to destabilize the stripes.  The regime where the pairing is strongest is broad,
and includes the case where the static stripes are completely absent
as well as the case where weak, smeared-out stripes are still
present.  In this regime
there is no evidence of stripe fluctuations in the density-density
correlation function; rather, the stripes
have completely or nearly evaporated into pairs. 

Our conclusions differ from those of previous approaches in
several respects.
Contrary to the frustrated phase separation
scenario, we do not have global phase separation; stripe formation is not
driven by competition with long range coulomb interactions;
and stripes and pairing compete, although there is a region in which both
coexist. In this coexistence regime, the pairing correlations
are two-dimensional rather than one-dimensional.
A key difference between the view we have
presented and the Hartree-Fock approach is that the local structure of the
domain wall we have discussed involves short-range
antiferromagnetic singlet bond correlations across the holes
rather than the mean-field,
single-particle correlations of the Hartree-Fock solution.
Furthermore, we find domain walls with a linear filling of
$\rho_\ell=0.5$ rather than $\rho_\ell=1$.

\acknowledgments

S.R.~White acknowledges support from the NSF under
grant \#DMR98-70930 and D.J.~Scalapino acknowledges support
from the NSF
under grant \#DMR95-27304.


%
\begin{figure}[ht]
\epsfxsize=2.5 in\centerline{\epsffile{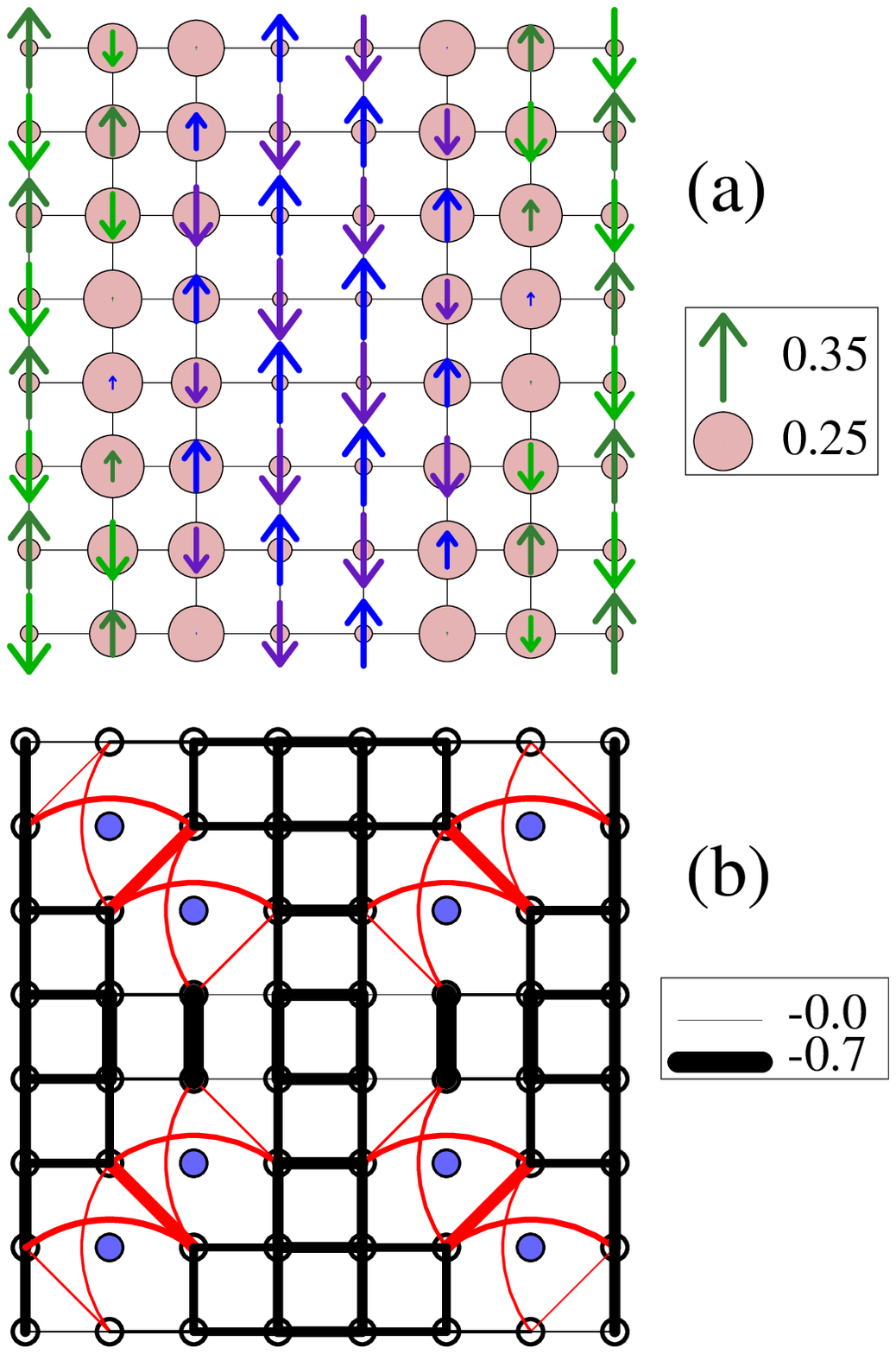}}
\caption{
(a) Hole density and spin moments on a center section of a $16\times 8$  $t-J$ 
lattice with $J/t=0.35$ and an average hole density $x=0.125$.
The diameter of the circles
is proportional to the hole density $1-\langle n_i\rangle$ on
the $i^{\rm th}$ site and the length of the arrows is
proportional to $\langle S^z_i\rangle$, according to the scales
shown. The arrows are color coded to show different
antiferromagetic domains. This structure
depends on the boundary conditions as discussed in the text.
(b) For an $8\times8$ system with $x=0.125$ whose hole density
and spin pattern are almost identical to that shown in (a), 
the blue dots show the most probable configuration of all the
holes and the strength of the exchange field $| \langle \vec S_i
\cdot \vec S_j\rangle |$ between two sites is denoted by the 
thickness of the line connecting the sites. 
Anomalous antiferromagnetic correlations
across holes are colored red. Only correlations where 
$ \langle \vec S_i\cdot \vec S_j\rangle < 0$ are shown.
}
\end{figure}

\begin{figure}[ht]
\epsfxsize=4.5 in\centerline{\epsffile{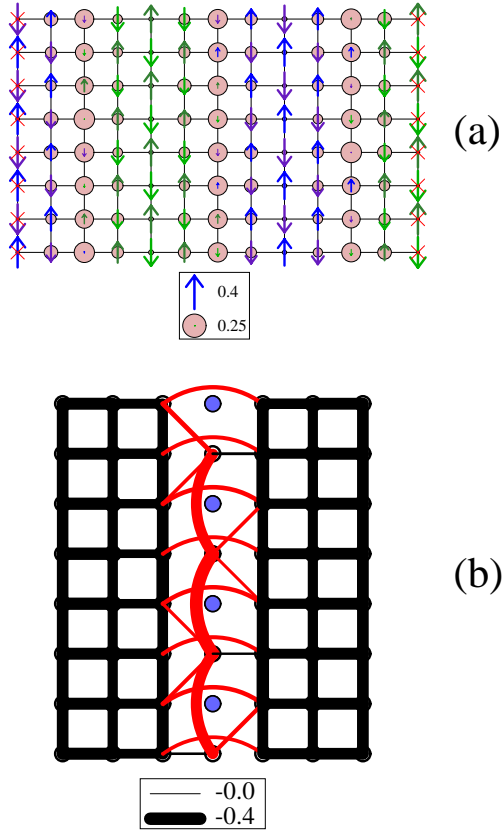}}
\caption{
(a) Hole density and spin moments on a $13\times 8$
lattice with cylindrical BCs, $J/t=0.35$ and $\pi$-shifted
staggered field $h$ on the open ends of magnitude $0.1$.
Here, there are 12 holes and the notation is similar to Fig.~1.
(b) For a $7\times8$ system with four holes and one domain wall, 
the most probable configuration of all the
holes and the strength of the exchange field surrounding them
is shown, with the same notation as in Fig. 1(b). 
This site-centered domain wall's hole density and spin patterns 
are nearly identical to those shown in (a).
Anomalous antiferromagnetic correlations
across holes are colored red. 
}
\end{figure}

\begin{figure}[ht]
\epsfxsize=3.0 in\centerline{\epsffile{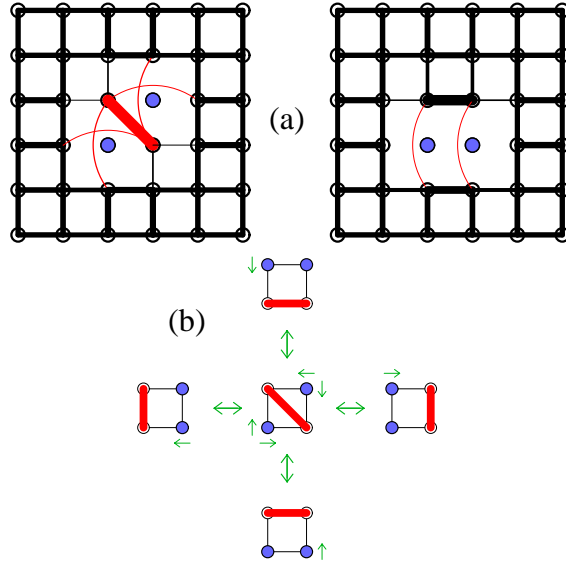}}
\caption{(a) The two most probable hole configurations and
associated local exchange field $\langle \vec S_i \cdot \vec S_j\rangle$
correlations in the
two-hole ground state of the system shown in Fig.~1(b) .
The scale and conventions are the same as in Fig.~1(b), but
only the central $6\times6$ region is shown. (b) Schematic
illustration showing the one-electron hops from the diagonal
configurations which lead to low energy configurations and
contribute to a lowering of the kinetic
energy.
}
\end{figure}
 
\begin{figure}[ht]
\epsfxsize=3.0 in\centerline{\epsffile{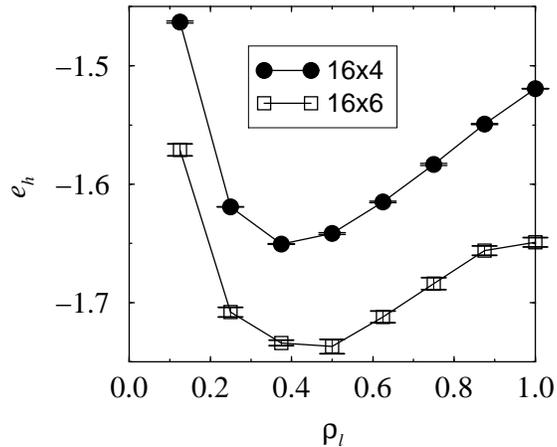}}
\caption{Energy per hole for a domain wall on $16\times4$ and
$16\times 6$ lattices with $J/t=0.35$ and the boundary 
conditions discussed in the text.
}
\end{figure}
 
\begin{figure}[ht]
\epsfxsize=3.0 in\centerline{\epsffile{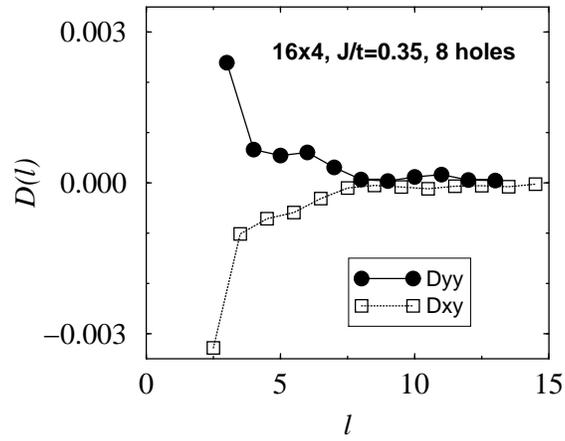}}
\caption{Pair field correlations along a domain wall running the length of
a $16\times 4$ lattice with $J/t=0.35$ and 8 holes.
}
\end{figure}

\begin{figure}[ht]
\epsfxsize=3.0 in\centerline{\epsffile{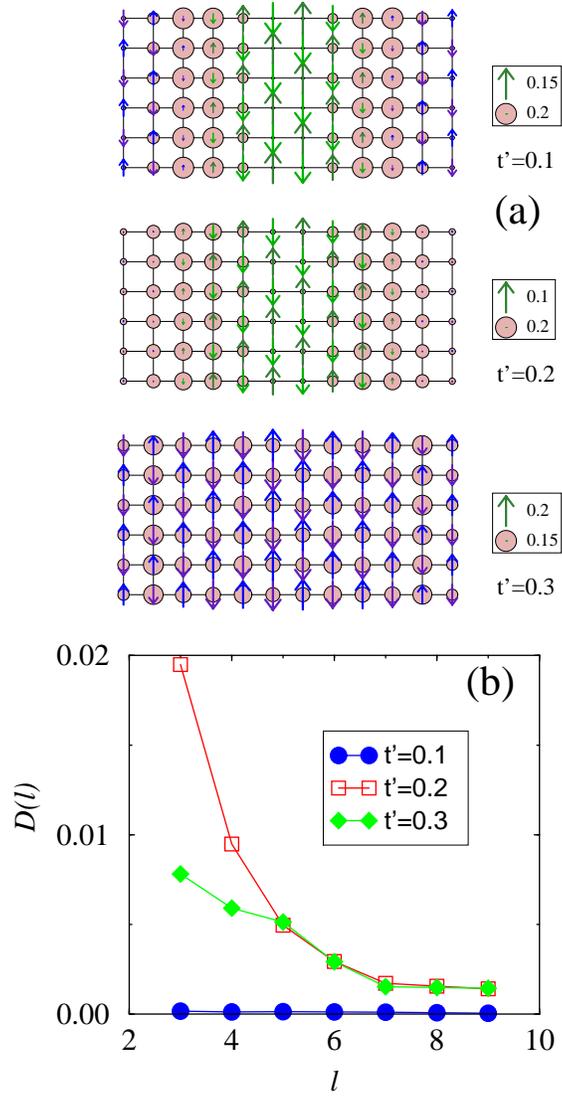}}
\caption{(a) Hole densities and spin moments on $12\times 6$ systems with
cylindrical boundary conditions.
(b) $d$-wave pairing correlations for the same system with $t^\prime=0.1$,
0.2, and 0.3.
}
\end{figure}

\end{document}